\documentstyle{article}
\begin{document}
\begin{center}
{\bf Nonperturbative and perturbative treatments of parametric
heating
in atom traps}\\
R. J\'auregui\\
{\it Instituto de F\'{\i}sica, Universidad Nacional Aut\'onoma de
M\'exico, Apdo. Postal 20-364, M\'exico, 01000,M\'exico}\\
{\it E-mail: rocio@fenix.fisica.unam.mx}
\end{center}
\begin{center}
{\bf Abstract}
\end{center}

We study  the quantum description of parametric heating in
harmonic potentials both nonperturbatively and perturbatively,
having in mind atom traps. The first approach establishes an
explicit connection between classical and quantum descriptions;
it also gives analytic expressions for properties such as the
width of fractional frequency parametric resonances. The second
approach gives an alternative insight into the problem and can be
directly extended to take into account nonlinear effects. This is
specially important for shallow traps.

\newpage

\section{Introduction}
Cooling techniques have allowed the trapping and manipulation of
atoms by optical means. Such systems are used to perform
experimental tests of fundamental principles and have important
applications, such as very precise frequency standards or studies
of cold atomic collisions or collective effects. Fluctuations of the
electromagnetic fields used to trap or to modify the quantum
state lead to decoherence in ion traps \cite{wineland} and limit
the trap stability in, $e.$ $g.$, far--of--resonance optical traps
\cite{savard}. As a consequence, there has been, in recent years,
an increasing interest in understanding the dynamics of
noise-induced heating in atom traps.

 In the harmonic model of the potential, fluctuations
 manifest themselves as either variations
 on the spring constant or on the equilibrium position. First order
 perturbative studies \cite{savard,gehm98}
 of harmonic parametric heating yield similar
 results to those classically expected: position noise is resonant
 at the vibrational frequency $\omega_0$ leading to a constant heating
 rate, while intensity fluctuations are  resonant at twice the
 vibrational frequency $2\omega_0$ leading to an exponential growth of the
 energy. Far-off-resonance optical traps
 are so sensitive to these fluctuations that parametric
 excitation has been used to accurately measure the trap
 parameters \cite{friebel98,vuletic98,roati}.

 There are other interesting phenomena predicted by
 the classical theory of harmonic parametric excitation such as
 resonance effects at fractional frequencies \cite{landau}.
 They arise from intensity modulations of frequency $2\omega_0/n$
 with $n$ being any integer number. The $n=2$ resonance has actually been
 observed in FORT's \cite{friebel98,vuletic98,roati}
 and resonances with $n\le 10$ have been classically studied in ion
 traps \cite{razvi}.

 The purpose of this paper is to analyze  the quantum
 description of parametric heating in harmonic potentials both
 nonperturbatively and perturbatively. The first treatment shows an
 explicit connection with the classical problem which is valid
 only for harmonic potentials. It is based on well known
 algebraic techniques \cite{gilmore}.It explains observed features of
 parametric heating predicted both classically and quantum
 mechanically. The second approach has the advantage that it can be
 directly extended to anharmonic potentials. This is specially important
 for shallow traps.

\section{Nonperturbative treatment of harmonic parametric heating}
\subsection{Time dependent equilibrium position}

The effective Hamiltonian that describes a harmonic oscillator
with fluctuations in the trap equilibrium position is
\begin{equation}
H=\frac{p^2}{2m} + \frac{1}{2}M\omega_0^2(q+\epsilon_q(t))^2
\end{equation}
where $\epsilon_q(t)$ measures such a fluctuation. When the
standard creation $a^\dagger$ and annihilation $a$ operators are
introduced, so that
\begin{eqnarray}
q &=& \sqrt{\frac{\hbar}{2m\omega_0}}(a + a^\dagger) \nonumber\\
p &=& -i\sqrt{\frac{\hbar m\omega_0}{2}} (a - a^\dagger)\quad,
\end{eqnarray}
this Hamiltonian can be written in the form
\begin{equation}
H = \hbar \omega_0 (a^\dagger a +1/2) + \hbar f_q(t)(a+a^\dagger)
+ \hbar g_q(t)
\end{equation}
with
\begin{equation}
f_q(t) = \sqrt{\frac{m\omega_0^3}{2\hbar}}\epsilon_q(t) \quad ;
\quad g_q(t) = \frac{m\omega_0^2}{2\hbar}\epsilon^2_q(t).
\end{equation}
The evolution operator $U(t)$ satisfies the equation
\begin{equation}
i\hbar \frac{\partial U}{\partial t} = H U
\end{equation}
with the boundary condition
\begin{equation}
U(0) = 1 . \label{boundary:1}
\end{equation}
Due to the fact that $\{a^\dagger a, a^\dagger, a, 1\}$ form a
closed algebra, it is reasonable to write \cite{gilmore}
\begin{equation}
U = e^{-i\lambda_1 a^\dagger}e^{-i\lambda_2 a}e^{\-i\lambda_3
a^\dagger a} e^{-i\lambda_4}.
\end{equation}
Using the well known relation
\begin{equation}
e^{\lambda A}Be^{-\lambda A} = B +\frac{\lambda}{1!} [A,B] +
\frac{\lambda^2}{2!}[A,[A,B]] + \frac{\lambda^3}{3!}[A,[A,[A,B]]]
+ \cdot\cdot\cdot \label{conm:1}
\end{equation}
 it can be directly shown that the $\lambda_i$'s satisfy the equations
 \begin{eqnarray}
 \dot \lambda_1 &=& -i\dot\lambda_3 \lambda_1 + f_q(t) \nonumber\\
 \dot\lambda_2 &=& i\dot\lambda_3\lambda_2 + f_q(t)
 \nonumber \\
 \dot \lambda_3 &=& \omega_0 \nonumber \\
 \dot\lambda_4 &=& \dot\lambda_3\lambda_2\lambda_1-i\dot\lambda_2\lambda_1 + g_q(t).
 \end{eqnarray}
 with the solution
  \begin{eqnarray}
 \lambda_1(t) &=& e^{-i\omega_0 t}\int_0^t f_q(t^\prime)e^{i\omega_0 t^\prime}
 dt^\prime \nonumber \\
\lambda_2(t) &=& e^{i\omega_0 t}\int_0^t
f_q(t^\prime)e^{-i\omega_0 t^\prime}
 dt^\prime  = \lambda_1^*(t)\nonumber \\
 \lambda_3(t) &=& \omega_0 t \nonumber\\
 \lambda_4(t) &=& \int_0^t(\omega_0\lambda_2\lambda_1 +i
 \dot\lambda_2(t^\prime)\lambda_1(t^\prime) +
 g_q(t^\prime))dt^\prime,
\end{eqnarray}
which guarantee the boundary condition (\ref{boundary:1}).
We are able, therefore, to evaluate the time evolution of any physical quantity.
For instance, it turns out that the energy operator evolves as
\begin{eqnarray}
\hat{\cal E}(t)&=&\hbar\omega_0
U^{-1}(t)(a^\dagger a + 1/2)U(t) \nonumber \\
&=&\hat{\cal E}(0) +\hbar\omega_0[ i\lambda_2 a-i\lambda_1
a^\dagger + \lambda_2\lambda_1)] . \label{energy:1}
\end{eqnarray}
If $\epsilon_q(t)$ is a fluctuating field, we are interested in
averages over a time T short compared with the time scale over which
the measurable physical quantities change, but large compared with
the correlation time of the fluctuations. Let us consider
fluctuations with zero mean value
\begin{equation}
\langle \epsilon_q(t)\rangle = \frac{1}{T} \int_0^T dt^\prime
\epsilon_q(t^\prime)=0,\label{m:1}
\end{equation}
and with a stationary correlation function
\begin{equation}
\langle \epsilon_q(t)\epsilon(t+\sigma)\rangle = \frac{1}{T}
\int_0^T dt^\prime
\epsilon_q(t^\prime)\epsilon_q(t^\prime+\sigma)=\eta_q(\vert
\sigma\vert).\label{m:2}
\end{equation}
Then
\begin{equation}
\langle \hat {\cal E}(t)\rangle = \hat{\cal E} (0) +
\frac{m\omega_0^4}{2}\int_0^tdt^\prime \int_0^t dt^{\prime\prime}
\langle\epsilon_q(t^\prime)\epsilon_q(t^{\prime\prime})\rangle
e^{i\omega_0(t^\prime - t^{\prime\prime})}. \label{james}
\end{equation}
When the one-sided spectrum of the position fluctuations in the
trap equilibrium position,
\begin{equation}
S_q(\omega) = \frac{2}{\pi}\int_0^\infty d\sigma
\cos(\omega\sigma) \eta_q(\vert \sigma\vert),
\end{equation}
is introduced and taking $t>>T$, it follows:
\begin{equation}
\langle {\cal E}(t)\rangle = {\cal E}(0) +
\frac{\pi}{2}m\omega_0^4S_q(\omega_0),
\end{equation}
recovering the asymptotic expression found by Savard et at
\cite{savard}. However, we have obtained it as valid in the
nonperturbative regime with only the assumption that the
fluctuating fields satisfy the equations (\ref{m:1}-\ref{m:2}).
 Notice that Eq.~(\ref{james}) is also valid whenever
 the state of the system is a Fock state. This possibility has been
 studied in Ref.~\cite{james} for the vacuum state where a detailed
 analysis of the short time behavior for different expressions of
 the correlation function $\lambda_1\lambda_2$ is performed.

 On the other hand, if one considers driven fluctuations
with a well specified $\epsilon_q(t)$ instead of noise fields, the
exact equation for the energy evolution (\ref{energy:1}) can be
used. For instance, if
 \begin{equation}
 \epsilon_q(t) = \epsilon_0\cos\omega_q t,
 \end{equation}
 $\lambda_1$ and $\lambda _2$ are trivially calculated
  and the energy of, say, a coherent state $\vert\alpha\rangle$
  evolves as
\begin{eqnarray}
\langle\alpha\vert \hat{\cal E}(t)\vert \alpha\rangle =
\hbar\omega_0 (\vert \alpha\vert^2 + \frac{1}{2})
&+&\frac{t}{2}\hbar\omega_0 \epsilon_0[i\alpha
(\zeta((\omega_q-\omega_0)t) +\zeta^*((\omega_q+\omega_0)t))-
\nonumber \\
&-&i\alpha^*(
 \zeta^*((\omega_q-\omega_0)t)+\zeta((\omega_q+\omega_0)t))]+\nonumber\\
&+&\frac{t^2}{4} \hbar\omega_0
\epsilon_0^2[\vert\zeta((\omega_q+\omega_0)t)\vert^2 +
\vert\zeta((\omega_q - \omega_0)t)\vert^2 +\nonumber \\&+&
2\cos(\omega_0t)\frac{\cos(\omega_0t)-\cos(\omega_qt)}{\omega_0^2
- \omega_q^2}]  \label{energy:2}
\end{eqnarray}
where we have defined
\begin{equation}
\zeta(y) = e^{iy/2}\frac{\sin(y/2)}{y/2}.
 \end{equation}
 Equation (\ref{energy:2}) shows the expected resonances at
$\omega=\omega_q$ and emphasizes the
relevance of the parameter $\epsilon_0 \alpha$ for characterizing
the expected heating rates.

\subsection{ Time dependent frequency}
In this case the effective Hamiltonian is
\begin{equation}
H=\frac{p^2}{2m} + \frac{1}{2}M\omega_0^2(1-\epsilon(t))q^2,
\end{equation}
which takes the form
\begin{equation}
H = \hbar\omega_o (a^\dagger a +1/2)
-\frac{\hbar\omega_0\epsilon(t)}{4}(a+a^\dagger)^2.
\end{equation}
Analogously to the former case, the closed algebra nature of $\{
a^\dagger a+1/2, a^{\dagger 2},a^2\}$ guarantees that the
evolution operator can be written in the form
\begin{equation}
U(t) = e^{c_0(a^\dagger a +1/2)/2}e^{c_- a^{\dagger 2}/2}e^{c_+
a^2/2}, \label{u:2}
\end{equation}
whenever the differential equations
\begin{eqnarray}
i\dot c_0 -2c_-\dot c_+ &=& 2\omega_0(1-1/2\epsilon(t))
\nonumber \\
i(\dot c_- - 2c_-^2\dot c_+)e^{c_0} &=&
\frac{1}{2}\omega_0\epsilon(t)
\nonumber \\
i\dot c_+ e^{-c_0} &=& \frac{1}{2}\omega_0\epsilon(t).
\label{nonlinear}
\end{eqnarray}
are satisfied together with the boundary conditions
\begin{equation}
c_0(0)=c_-(0)=c_+(0) =0.
\end{equation}
 A connection between the functions $\{c_0,c_+,c_-\}$ and
 a pair of solutions $h_1(t)$ and $h_2(t)$ of the
 classical equations of motion
\begin{equation}
\ddot h_k(t) + \omega_0^2(1 -\epsilon(t))h_k(t)=0
\label{hk:1}
\end{equation}
with the boundary conditions
\begin{eqnarray}
h_1(0)=1\quad &,&\quad \dot h_1(0) =0 \nonumber\\
h_2(0)=0\quad &,&\quad \dot h_2(0) =-\omega_0. \label{b2:2}
\end{eqnarray}
is explicitly given by \cite{casta97}:
\begin{eqnarray}
c_0 &=& -2ln(M_1)\nonumber \\
c_+ &=& M_2^*/M_1 \nonumber \\
c_- &=& -M_2M_1 \label{c:2}
\end{eqnarray}
where
\begin{eqnarray}
M_1 =\frac{1}{2}[(h_1 -ih_2) -\frac{1}{\omega_0}(\dot h_2 + i
\dot h_1)]\nonumber \\
M_2 =\frac{1}{2}[-(h_1 -ih_2) -\frac{1}{\omega_0}(\dot h_2 + i
\dot h_1)].\label{mi:2}
\end{eqnarray}
Eqs.~(\ref{u:2},\ref{hk:1}-\ref{mi:2}) show that the theory
of classical response of harmonic oscillators  to noise
 can be useful in the study of their quantum dynamics.

Using Eq.(\ref{conm:1}), an expression for the time evolution of
the energy is found:
\begin{eqnarray}
\hat{\cal E}(t) &=& \hbar\omega_0 U^\dagger(t)(a^\dagger a
+1/2)U(t)
\nonumber \\
&=& \hbar\omega_0(1+2\vert M_2\vert^2)[(a^\dagger a+1/2)
-M_1M_2a^2-M_1^*M_2^*a^{\dagger 2}],
\label{energy:3}
\end{eqnarray}
as well as a nonperturbative expression for nonzero transition
probabilities between Fock states $\vert k\rangle$ and $\vert s
\rangle$:
\begin{equation}
\vert \langle k\vert U\vert s\rangle\vert^2
 = \frac{\vert
M_2\vert^{k-s}}{2^{k-s}\vert M_1\vert^{k+s+1}} \big[
\sum_{m=0}^{[s/2]}(-1)^m\frac{\sqrt{s!k!}}{m!(s-2m)!(m+(k-s)/2)!}
\big(\frac{\vert M_2\vert}{2}\big)^{2m}\big]^2 \label{nonpert:tp}
\end{equation}
where  $s$ and $k$ have the same parity and for definiteness we
have taken $k\ge s$.

Let us focus in the particular case
\begin{equation}
\epsilon(t) = \epsilon_0 \cos \omega t.
\end{equation}
corresponding to controlled parametric excitation.
 Then, the classical equations of motion can be
written in the Mathieu canonical form
\begin{equation}
\frac{d^2 h_k}{d^2z} +(\alpha-2q\cos 2z)h_k =0,
\end{equation}
with $z=\omega t/2$, $\alpha =(2\omega_0/\omega)^2$ and $q
=\epsilon_0 \alpha/2$. It is well known \cite{abramo,nonlinear}
that depending on the values of $\alpha$ and $q$, the solutions
$h_k$ are stable or unstable. In the context of parametric
heating the transition curves separating regions of stability and
instability define the width of the corresponding resonance. For
$q<<\alpha$ the resonances are located at $\alpha \sim n^2$, $i.$
$e.$, $\omega =2\omega_0/n$. Their width can be found as a power
series in $\epsilon_0$ using, $e.$ $g.$, equations $20.2.25$ of
Ref.~\cite{abramo}. Thus, the resonance at $\omega=2\omega_0$ has
a width $\Delta_{2\omega_0}\sim \epsilon_0\omega_0/2$ while the
resonance $\omega=\omega_0$ has a width $\Delta_{\omega_0}\sim
\epsilon_0^2\omega_0/6$.

Using the normal form of the unstable solutions of the classical
equations of motion, it is found that
\begin{equation}
M_i = \mu^{(+)}_i e^{\gamma \omega t/2}\phi_0(t) +\mu^{(-)}_i
e^{-\gamma \omega_0 t/2}\phi_0(-t), \label{phit}
\end{equation}
with $\phi_0$ a periodic function $\phi_0(t) =
\phi_0(t+2\pi/\omega)$, and $\gamma =\gamma_r +i\gamma_i$ a
complex number known as the characteristic exponent. The complex
numbers $\mu^{\pm}_i$ are determined by the boundary conditions
(\ref{b2:2}) and the general form of $M_i$ given by Eq.~(\ref{mi:2}).
Thus,  for $\vert \gamma_r\omega \vert t
>>1$ the energy (\ref{energy:3}) in the resonance region behaves
as
\begin{eqnarray}
\hat{\cal E}(t) &\rightarrow &
\hbar\omega_0[(1+2\vert\mu^{(+)}_3\vert^2 e^{\vert \gamma_r\vert
\omega t})(a^\dagger a +1/2)+\nonumber\\ &-&
\mu^{(+)}_2\mu^{(+)}_1 e^{\vert\gamma_r\vert\omega t}
a^{\dagger^2} -\mu^{(+)*}_2\mu^{(+)^*}_1
e^{\vert\gamma_r\vert\omega t} a^2].
\end{eqnarray}
If the initial state is a Fock state, the energy will exhibit an
exponential growth with a rate determined by the characteristic
exponent.

In order to illustrate these ideas, the time evolution
of the transition probability between the vacuum and the
second excited state is shown in Fig.~(1) for $\epsilon_0 = 0.05$, and
$\omega= \omega_0$, $i.$ $e.$, the first fractional resonance. We
observe that for $\omega t \le 1$ the transition probability
exhibits an  approximately polynomial growth. However, for longer
times an oscillatory behavior arises. In Fig.~(2), we illustrate
the energy evolution starting from vacuum state; we take the same
values for the parameters $\omega$ and $\epsilon_0$. An
exponential growth superposed to the oscillatory behavior of
function $\phi(t)$, in Eq.~(\ref{phit}), is found. The  similarities
between Fig.~(1b) and Fig.~(2) are due to the fact that the
main source of heating from vacuum state are precisely the transitions
$\vert 0\rangle \rightarrow \vert 2 \rangle$. These graphs have
been obtained using a numerical solution of the Mathieu equation.

\section{Perturbative approach to parametric heating.}
The nonperturbative approach to parametric heating made in the last
section was useful to understand the connection between classical
and quantum descriptions of heating, due to  either controlled or
stochastic variations of the parameters defining a harmonic
oscillator. Unfortunately, this approach is valid only for
quadratic potentials while nonlinear effects may be determinant
 in experiments with shallow confining potentials
\cite{razvi,jauregui01}. The purpose of this section is to study
some high order perturbative effects due to variations on the
strength of a confining potential. The resulting equations will
be applied to a harmonic oscillator but with a straightforward extension
to anharmonic potentials. Besides, this will allow us to
understand fractional frequency resonances from an alternative
point of view.

 The system is described by a Hamiltonian
\begin{equation}
H = \frac{p^2}{2m} +V(x)(1 +\epsilon_v(t))\quad \epsilon_v(t)<<1.
\end{equation}
Following standard time dependent perturbation theory we define
the unperturbed Hamiltonian
\begin{equation}
H_0 = \frac{p^2}{2m} + V(x)
\end{equation}
and work in the interaction picture in which the equation of
motion of the state is
\begin{equation}
i\hbar \frac{d\vert \tilde\Psi(t)\rangle}{dt} =
\epsilon_v(t)\tilde V(x,t) \vert \tilde\Psi(t)\rangle.
\end{equation}
The transformed state $\vert \tilde \Psi(t)\rangle$ in the
interaction picture is obtained from the Schr\"odinger picture
state vector by a time-dependent unitary operator
\begin{equation}
\vert\tilde\Psi(t)\rangle = e^{iH_0t/\hbar}\vert\Psi(t)\rangle,
\end{equation}
while the interaction operator $\tilde V(x)$ is given by
\begin{equation}
\tilde V(x,t) = e^{iH_0t/\hbar}V(x)e^{-iH_0t/\hbar}.
\end{equation}
In this picture the evolution operator satisfies the integral
equation
\begin{equation}
\tilde U(t) = 1 -\frac{i}{\hbar}\int_0^t \epsilon_v(t)\tilde
V(x,t) \tilde U(t^\prime)dt^\prime. \label{u:1}
\end{equation}

Let us consider the transition probability amplitude $\langle
k\vert U (t)\vert s\rangle$ between given eigenstates of the
unperturbed Hamiltonian
\begin{equation}
H_0\vert n\rangle = E_n\vert n \rangle.
\end{equation}
An iterative treatment of Eq.~(\ref{u:1}) gives
\begin{eqnarray}
\langle k\vert \tilde U(t)\vert s\rangle &=& \delta_{ks} -
\frac{i}{\hbar}V_{ks}\int_0^t dt^\prime \epsilon_v(t^\prime)
e^{i\omega_{ks}t^\prime}\nonumber\\
&-&\frac{1}{\hbar^2} \sum_n V_{kn}V_{ns}\int_0^tdt^\prime
\epsilon_v(t^\prime) e^{i\omega_{kn}t^\prime} \int_0^{t^\prime}
dt^{\prime\prime} \epsilon_v(t^{\prime\prime})
e^{i\omega_{ns}t^{\prime\prime}}+ \cdot\cdot\cdot
\end{eqnarray}
with $\omega_{kn} = (E_k - E_n)/\hbar$ and
\begin{equation}
V_{kn}=:\langle k\vert V\vert n\rangle = E_k \delta_{kn} - \langle
k\vert \frac{p^2}{2m}\vert n\rangle .
\end{equation}
If heating is induced by a controlled modulation of the confining
potential
\begin{equation}
\epsilon_v(t) = \epsilon_0 \cos \omega t,
\label{source:1}
\end{equation}
then up to second order in $\epsilon_0$:
 \begin{eqnarray}
\langle k\vert\tilde U^{(2)}(t)\vert s\rangle = \delta_{ks} +
\frac{i}{2\hbar}\epsilon_0 t V_{ks}[\zeta((\omega_{ks}+\omega)t)
&+&
\zeta((\omega_{ks}-\omega)t)]+\nonumber\\
-\big(\frac{\epsilon_0}{2\hbar}\big)^2t\sum_n V_{kn}V_{ns}
\big[\frac{1}{i(\omega_{ns}+\omega)}
\big(\zeta((\omega_{ks}+2\omega)t)&+&\zeta(\omega_{ks}t)-\nonumber\\
- \zeta((\omega_{kn}+\omega)t)
&-&\zeta((\omega_{kn}-\omega)t)\big)
+ \nonumber \\
 \frac{1}{i(\omega_{ns}-\omega)}
\big(\zeta((\omega_{ks}-2\omega)t)&+&\zeta(\omega_{ks}t)
-\nonumber\\ \zeta((\omega_{kn}+\omega)t)
&-&\zeta((\omega_{kn}-\omega)t)\big) \big]. \label{reso:1}
\end{eqnarray}
These expressions have physical meaning only if  the changes in
the wave function induced by $\tilde U(t)$ are small in the
interval $(0,t)$.

For a harmonic oscillator with frequency $\omega_0$, one finds that
\begin{equation}
V_{kn} = \frac{\hbar\omega_0}{4} [(2k+1)\delta_{kn}
+\sqrt{k(k-1)}\delta_{k,n+2} + \sqrt{(k+1)(k+2)}\delta_{k,n-2}]
\end{equation}
As a consequence, the following products $V_{kn}V_{ns}$ may be
different from zero:

(i) $V_{k,k\pm 2}V_{k\pm 2,k\pm 4}$. In (\ref{reso:1}), the
resonant terms appear in the combination $\zeta((4\omega_0\pm
2\omega)t))-\zeta((2\omega_0\pm \omega)t)$ so that this
transition is highly suppressed.

(ii) $V_{kn}V_{nk}$. Resonances are located at $\omega =
0,2\omega_0$. For $\omega\sim 2\omega_0$ the contribution of the
transition amplitude  is of the form
\begin{equation}
\frac{\epsilon^2t}{2\hbar\omega_0} \big[\vert V_{k,k+2}\vert^2
\big(\frac{\zeta(0) -
\zeta((\omega-2\omega_0)t)}{i\omega-\omega_0}\big) +\vert
V_{k,k-2}\vert^2 \big(\frac{\zeta(0) -
\zeta((2\omega_0-\omega)t)}{i\omega-\omega_0}\big)].
\end{equation}

(iii) $V_{k,k\pm 2}V_{k\pm 2,k\pm2}$ and $V_{k,k}V_{k,k\pm 2}$.
These are transitions that may be viewed as a combination of two
virtual transitions $k\rightarrow s\rightarrow s$ and
$k\rightarrow k\rightarrow s$. The corresponding resonance
frequency according to Eq.~(\ref{reso:1}) is the fractional
frequency $\omega =\vert \omega_{ks}\vert /2 = \omega_0$. In fact
the transition probability for a modulating frequency
$\omega\sim\omega_0$ is
\begin{eqnarray}
\vert\langle k\vert U^{(2)}\vert s\rangle\vert^2&\sim&
\big(\frac{\epsilon}{2\hbar}\big)^4\frac{t^2}{\omega_0^2} \vert
V_{ks}\vert^2
(V_{kk}-V_{ss})^2\frac{\sin^2(\omega-\omega_0)t/2}{(\omega-\omega_0)^2t^2/4}
\nonumber \\
&\sim& \frac{\epsilon^4\omega_0^2t^2}{1024}(k-s)^2[k(k-1)\delta_{k,s+2}
+(k+1)(k+2)\delta_{k,s-2}]\nonumber \\
&\cdot&\frac{\sin^2(\omega-\omega_0)t/2}{(\omega-\omega_0)^2t^2/4}
\label{compare:gra}
\end{eqnarray}

In all cases, the nonresonant terms $\zeta(\omega^\prime t)$,
with $\omega^\prime \ne 0$, give rise to an oscillatory behavior
of the transition probability which is consistent with that found
in Fig.~1 for the exact evolution of the transition probability
$\vert\langle 0\vert U\vert 2\rangle\vert^2$. If one considers
sufficiently long times, $\omega^\prime t >>1$, having in mind
the delta-function representation
\begin{equation}
\delta(\omega) = \frac{2}{\pi}\lim_{t\rightarrow \infty}
\frac{\sin^2(\omega t/2)}{t\omega^2},
\end{equation}
it is clear that just the resonant terms have a significant
contribution. In such a limit the transition probability rates
$R_{s\leftarrow k}$ are constant.

Now, some of the general behavior of higher order corrections can
be inferred. Thus, the dominant transition probability of a
fractional frequency resonance $\omega=2\omega_0/n$ arises at
n-order perturbation theory. It can be interpreted as an n-step
procedure consisting on n-virtual transitions , where $n-1$ of them do
not change the state and one changes it. Thus, we expect  the
expression (\ref{compare:gra}) to describe approximately the
transition probabilities $k\rightarrow k\pm2$ when the source has
a frequency $\omega=\omega_0$. This can be verified by comparing
with the exact results of last section.

\section{Discussion and Outlook.}

 In this work, we have performed a perturbative and a nonperturbative
analysis of quantum parametric oscillators. The first approach is
based on standard algebraic techniques and gives a direct
connection between classical and quantum results. In the case of
controlled driving terms of the form $\epsilon(t) = \epsilon_0
\cos\omega t$, the analytic solutions were used to evaluate
time-dependent observables such as the energy growth due to
parametric heating. This is specially important for
far--off--resonance--traps (FORT's) when the harmonic oscillator
approximation is valid. In that case, parametric heating is used
as a technique to measure the characteristic frequency of the
trap. In such experiments, the fractional frequency resonances
are usually observed and it is clear, within our formalism,
how to perform their quantum description. In particular, using well known
results of the theory of Mathieu functions, it was shown how to
evaluate the fractional resonances width and how  obtain
explicit expressions for the exponential growth of the energy.

The possibility of using the nonperturbative analysis to describe
noise heating effects is a subject that deserves more analysis.
This idea has recently been exploited \cite{james} in the case of
fluctuations in the trap center.  In the case of intensity
fluctuations, the problem is more complicated since a rigorous
description would require to study the correlation functions of
the classical solutions $\{h_1,h_2\}$ and their derivatives
$\{\dot h_1,\dot h_2\}$ as they appear in the expressions for the
time evolution of a given physical quantity. For instance, to
study the time-evolution of the energy, it would be necessary to
evaluate the classical correlation functions $\vert M_2\vert^2$
and $M_1M_2$. These correlation functions would depend on the
noise correlations.

An alternative would correspond to use solutions $\{h_1,h_2\}$ of
classical equations describing in an effective way the coupling
between  the harmonic oscillator and the fluctuating  fields. The
usefulness of such an approach would be conditioned by its ability
to reproduce experimental effects. For instance, classical
harmonic oscillators with an intensity variation $\epsilon(t) =
\epsilon_0\cos\omega^\prime t$ and subject to a damping force
$-\gamma \dot x$ lead to fractional frequency resonances only if
$\epsilon_0$ is greater than a certain threshold which
depends on $\gamma$ and the order of the resonance. Such
thresholds have been observed in the classical collective motion
of ions in Paul traps \cite{razvi}. It would be expected that a
similar phenomena occurs in the quantum regime of motion. The
experimental study of these thresholds could be used to evaluate
effective damping effects in atom traps.

In this paper, the perturbative analysis of heating induced by
variations of the intensity of a potential was studied in detail
for a harmonic oscillator. This analysis gave a different
insight into the problem.  Fractional frequencies of order n appear
in n-th order perturbation theory. From the quantum theory point
of view, they are a direct consequence of: (i)the fact that the harmonic
potential has  diagonal matrix elements $\langle k \vert V \vert
k\rangle$ different from  zero; (ii) the equidistant spectrum of
the harmonic oscillator.

The extension of the perturbative analysis to other confining
potentials is straightforward. In fact, this approach has already been
implemented for FORT's with shallow potentials \cite{jauregui01}.

We thank S. Hacyan, P. Nicola and G. Modugno for stimulating discussions.

{\bf Figure Captions}

Figure 1.

Nonperturbative time evolution of the transition
probability $\vert\langle 0\vert U\vert 2\rangle\vert^2$ due to
parametric heating with a time dependence $\epsilon(t) =
\epsilon_0 \cos\omega t$,
 when $\epsilon_0 = 0.05$ and
the first fractional resonance condition $\omega = \omega_0$ is satisfied.
The short time behavior is shown in Fig 1a.

Figure 2.

Nonperturbative evolution of the energy of a time dependent
harmonic oscillator of natural frequency $\omega_0$ in units
of $\hbar \omega_0$. The system is initially in the vacuum state
 and the intensity time dependence is given by
$\epsilon(t) = \epsilon_0 \cos\omega t$, $\epsilon_0 = 0.05$,
$\omega=\omega_0$, so that the first fractional resonance
condition is satisfied.  Superposed to the exponential growth, the
oscillatory behavior of the function $\phi(t)$ that appears in
the Floquet solution to Mathieu equation is clearly manifested.

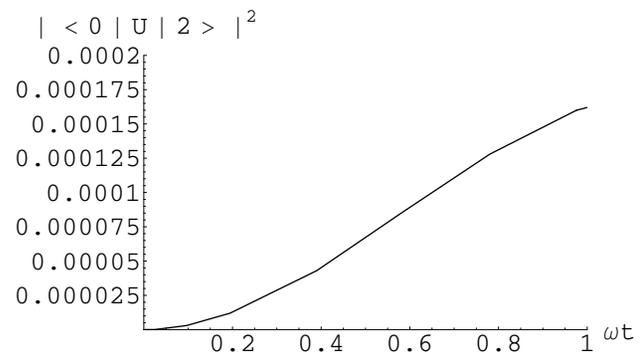
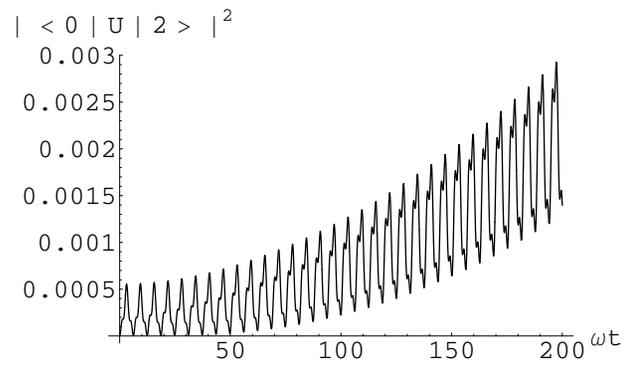
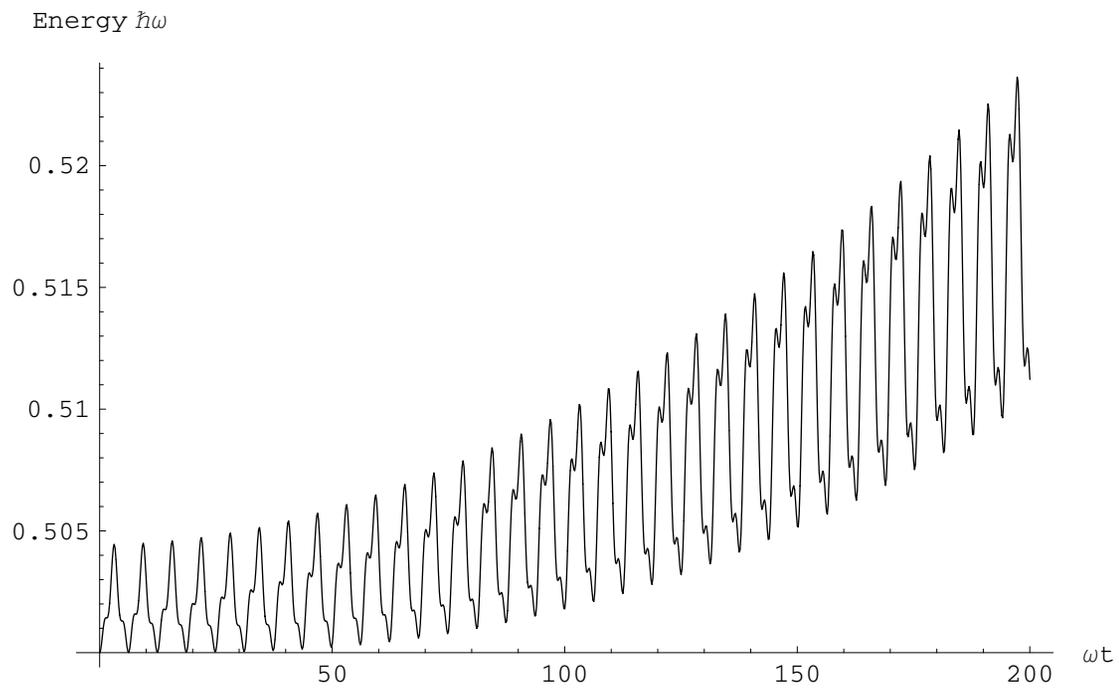
\end{document}